\documentclass[a4paper]{jpconf}
\usepackage{graphicx}
\usepackage{siunitx}
\usepackage{caption}
\usepackage{subcaption}
\usepackage{hyperref}
\usepackage{amsmath}

\begin{document}
\title{Development of a fast simulator for GEM-based neutron detectors}

\author{R. Felix dos Santos$^{1}$, M. G. Munhoz$^1$, M. Moralles$^2$, L. A. Serra Filho$^1$, M. Bregant$^1$ and F. A. Souza$^2$}

\address{$^1$Instituto de Física da Universidade de São Paulo, Rua do Matão, 1371, Cidade Universitária,  São Paulo, Brasil}
\address{$^2$Instituto de Pesquisas Energéticas e Nucleares, Avenida Lineu Prestes, 2242, Cidade Universitária, São Paulo, Brasil}

\ead{renanfelix@usp.br}

\begin{abstract}
Gas Electron Multiplier (GEM)-based detectors using a layer of $^{10}\mathrm{B}$ as a neutron converter is becoming popular for thermal neutron detection. A common strategy to simulate this kind of detector is based on two frameworks: Geant4 and Garfield++. The first one provides the simulation of the nuclear interaction between neutrons and the $^{10}\mathrm{B}$ layer, while the second allows the simulation of the interaction of the reaction products with the detector gas leading to the ionization and excitation of the gas molecules. Given the high ionizing power of these nuclear reaction products, a full simulation is very time consuming and must be optimized to become viable. In this work, we present a strategy to develop a fast simulator based on these two frameworks that will allow us to generate enough data for a proper evaluation of the expected performance and optimization of this kind of detector. We will show the first results obtained with this tool concentrating on its validation and performance.
\end{abstract}

\section{Introduction}
Due to the shortage of helium-3, widely used in thermal neutron gaseous detectors, alternatives need to be studied to continue producing this kind of detector. Gas Electron Multipliers (GEM) \cite{Sauli1997} are a type of Micro-Pattern Gaseous Detectors (MPGD), widely used in particle tracking systems, as the Time Projection Chamber of the ALICE experiment in the LHC-CERN \cite{Adolfsson2021}, and proposed for many other applications, including neutron detection. Neutrons can be detected indirectly through a nuclear reaction where the products are ionizing radiation. In our application, we are using $^{10}\mathrm{B}$ as a neutron converter to induce the nuclear reaction:

\begin{equation}
^{10}\mathrm{B}+n \longrightarrow 
\begin{cases}
^{4}\mathrm{He} + ^{7}\mathrm{Li}: \quad \text{ 2.792 MeV (ground state, about 6\%)}\\
^{4}\mathrm{He} + ^{7}\mathrm{Li}^{*}: \quad \text{2.310 MeV (excited state, about 94\%).}
\end{cases}
\end{equation}

\section{Simulation Tools}
A common strategy to simulate this kind of detector is based on the frameworks: GEANT4 \cite{Agostinelli2003} and Garfield++ \cite{schindler}:
    \begin{itemize}
      \item \textbf{GEANT4} - Using the physics list \textit{QGSP\_BERT\_HP}, which has high precision models for low energy neutrons \cite{geant4_physicsReference}, we simulate the nuclear interaction of \SI{41.8}{\milli\eV} thermal neutrons with the boron layer, producing charged particles, as well as the transport of these particles inside the detector.
      \vspace{0.1cm}
      \item \textbf{Garfield++} - The electric field was calculated employing the finite element method through Elmer \cite{Ruokolainen2021}. The ionization pattern due to low energy ions were produced using SRIM \cite{Ziegler2010}, while the properties of electrons in the gas mixture was determined with Magboltz \cite{Biagi1999}.
    \end{itemize}

Given the high ionizing power of the nuclear reaction products from $^{10}\mathrm{B}(n,\alpha)^{7}\mathrm{Li}$ reaction, a full simulation is very time consuming, mainly due to Garfield++, and must be optimized to become viable. Thus, in the following, we present a strategy to develop a fast simulator based on these two frameworks.

\section{Parametrization Strategies} 
\begin{figure}[h]
      \centering
        \includegraphics[width=0.70\textwidth]{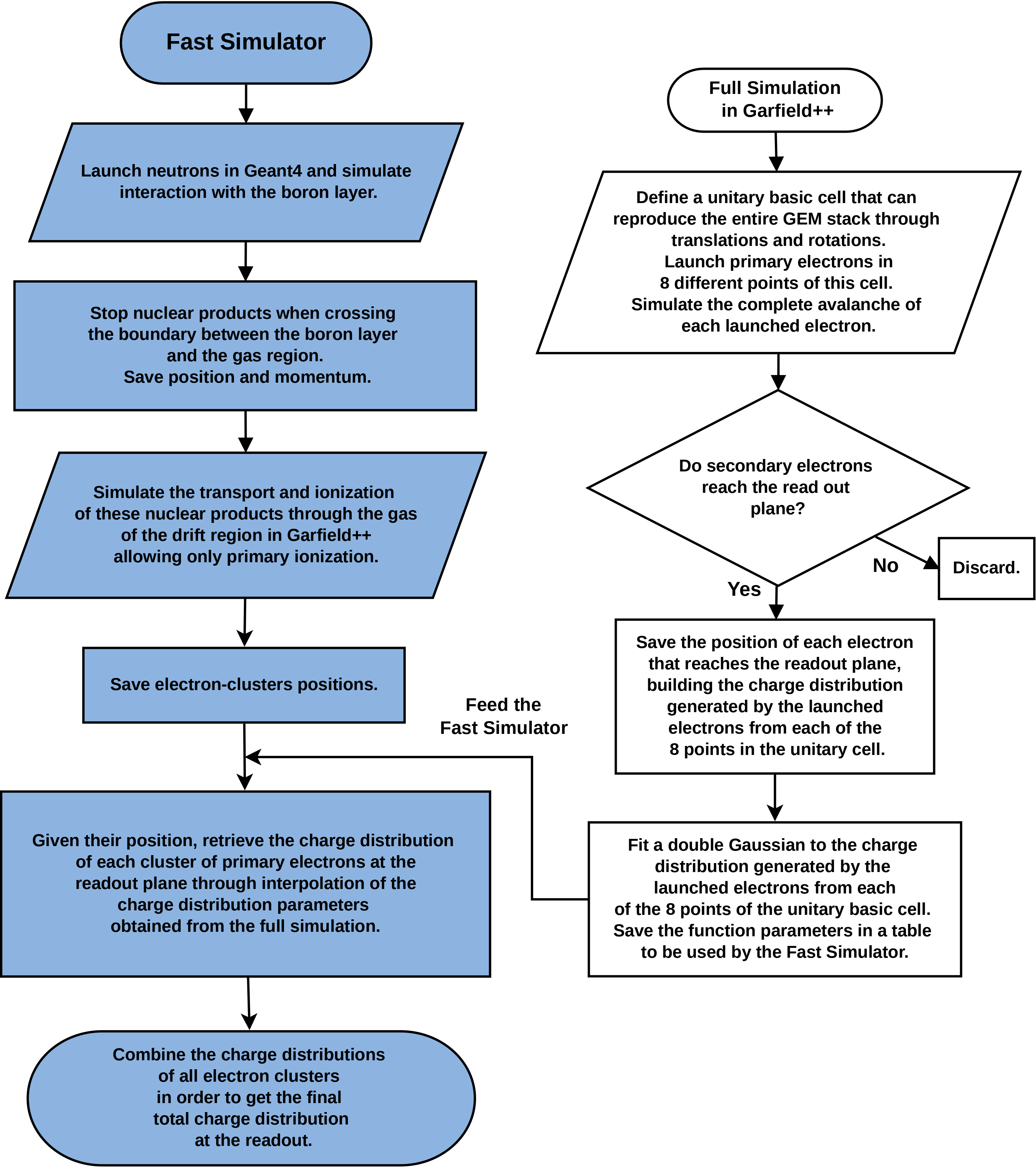}
        \caption{Fast simulator diagram flow.}
        \label{fig:diagram}
    \end{figure}

The fast simulator steps are represented in the diagram shown in Fig. \ref{fig:diagram}.

\begin{figure}[!htp]
      \centering
      \begin{subfigure}[c]{0.3\textwidth}
        \centering
        \includegraphics[width=\textwidth]{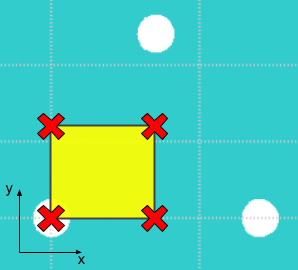}
        \caption{Unitary basic cell.}
        \label{fig:basicCell}
        \end{subfigure}
        \begin{subfigure}[c]{0.37\textwidth}
        \centering
        \includegraphics[width=\textwidth]{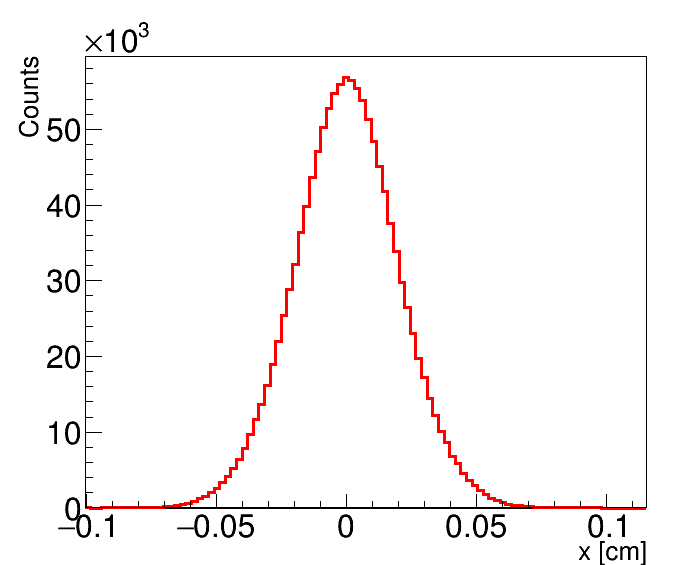}
        \caption{Example of the charge distribution at the readout.}
        \label{fig:SIMx}
       \end{subfigure}
       \caption{Parametrization strategies.}
    \end{figure}

Given the symmetry of the GEM foil, the entire detector can be represented by translations and rotations of a unitary basic cell, as represented in Fig. \ref{fig:basicCell}, where there are 3 white circles representing the GEM holes. Four points are marked with a red X, corresponding to the positions in the transverse (readout) plane where the primary electrons are launched in the full simulation in two different positions in the longitudinal (drift) direction. The rectangular area in yellow can be translated and rotated to map the entire GEM. The charge distribution at the readout, as shown in Fig. \ref{fig:SIMx}, is better fitted with a double Gaussian. And the primary ionization is allowed only in the drift region.

\section{Double-GEM Detector Prototype}
\begin{figure}[h]
      \centering
      \begin{subfigure}[c]{0.40\textwidth}
        \centering
        \includegraphics[width=\textwidth]{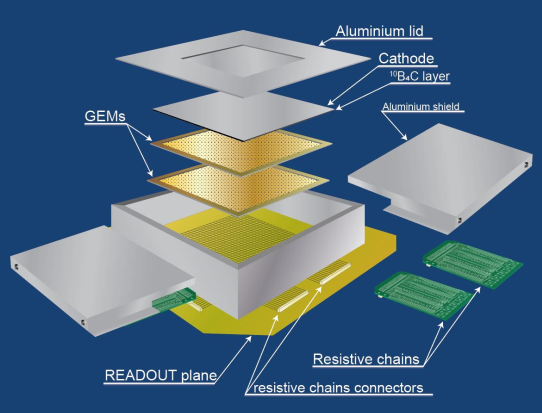}
        \caption{Double-GEM detector prototype.}
        \label{fig:prototype}
      \end{subfigure}
        \begin{subfigure}[c]{0.27\textwidth}
        \centering
        \includegraphics[width=\textwidth]{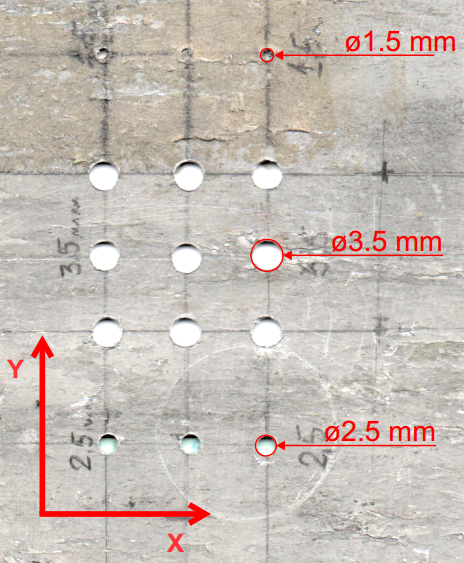}
        \caption{Cadmium mask.}
        \label{fig:maskCd}
       \end{subfigure}
       \caption{Experimental prototype.}
    \end{figure}

In order to validate the simulations, data from an experimental prototype, sketched in Fig. \ref{fig:prototype}, was used. It is a double-GEM composed of a stack of $\SI{0.5}{\milli\meter}$ thick aluminum cathode coated with enriched boron carbide and two GEM foils. The aluminum lid and the cathode are $\SI{3}{\centi\meter}$ apart from each other. The drift, transfer and induction regions were set to $\SI{2}{\milli\meter}$, $\SI{1}{\milli\meter}$ and $\SI{1}{\milli\meter}$ thick and bias of $\SI{100}{\volt}$, $\SI{300}{\volt}$ and $\SI{400}{\volt}$, respectively. It is operated with $\mathrm{Ar}/\mathrm{CO}_2$ (90/10) gas mixture.

\section{Preliminary Results}
\begin{figure}[h]
      \centering
      \begin{subfigure}[t]{0.325\textwidth}
        \centering
        \includegraphics[width=\textwidth]{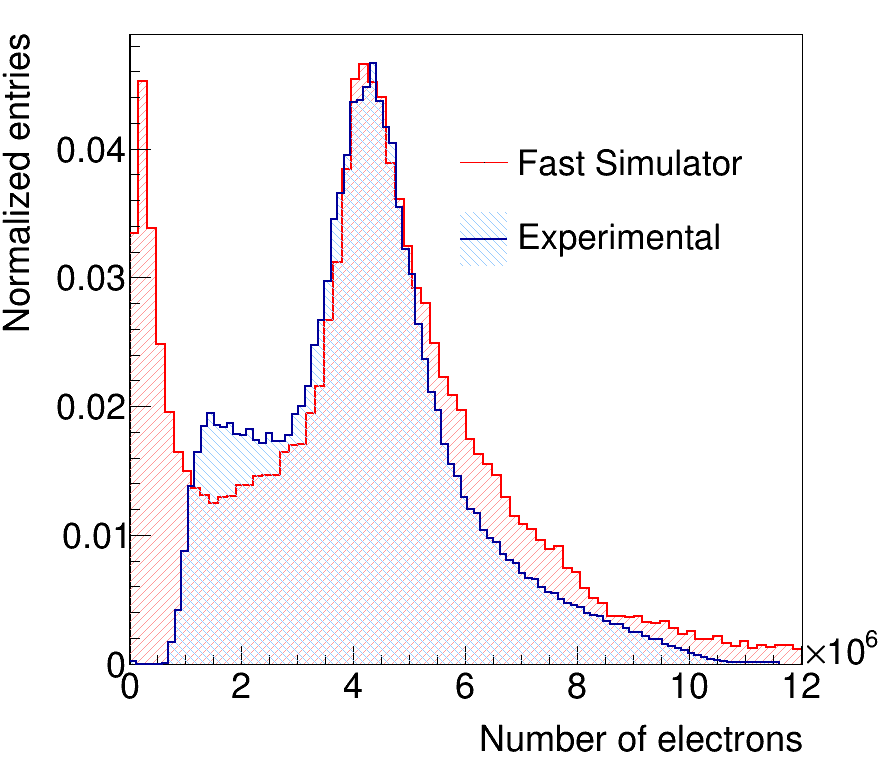}
        \caption{Spectrum of charge collected at the readout plane.}
        \label{fig:chargeDist}
      \end{subfigure}
      \hfill
    \begin{subfigure}[t]{0.325\textwidth}
        \centering
        \includegraphics[width=\textwidth]{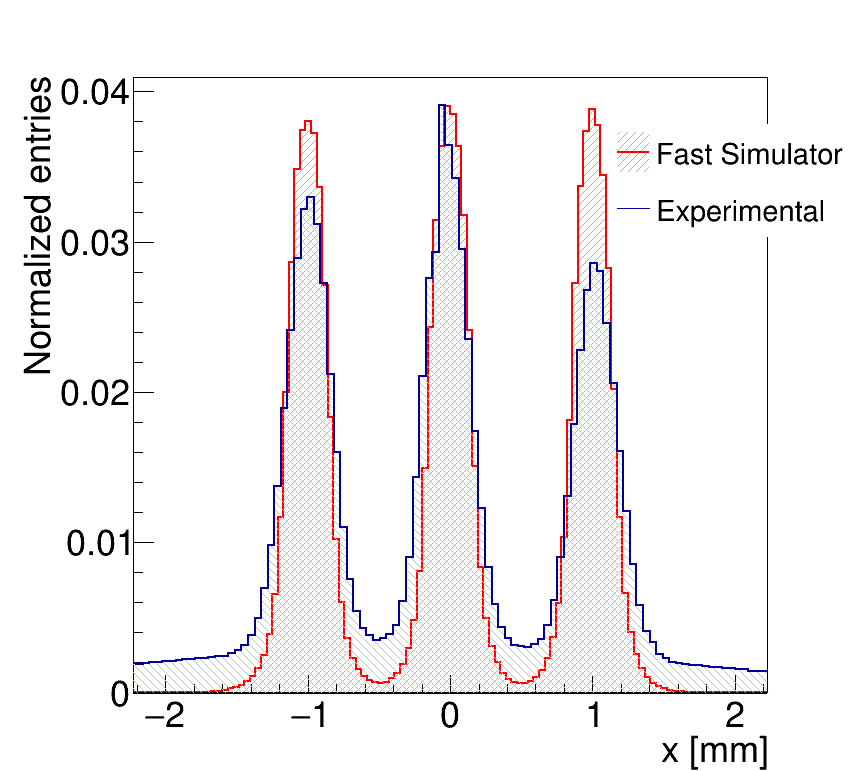}
        \caption{Projection in x-axis for the three central holes and same diameter.}
        \label{fig:proj_x}
    \end{subfigure}
    \begin{subfigure}[t]{0.325\textwidth}
        \centering
        \includegraphics[width=\textwidth]{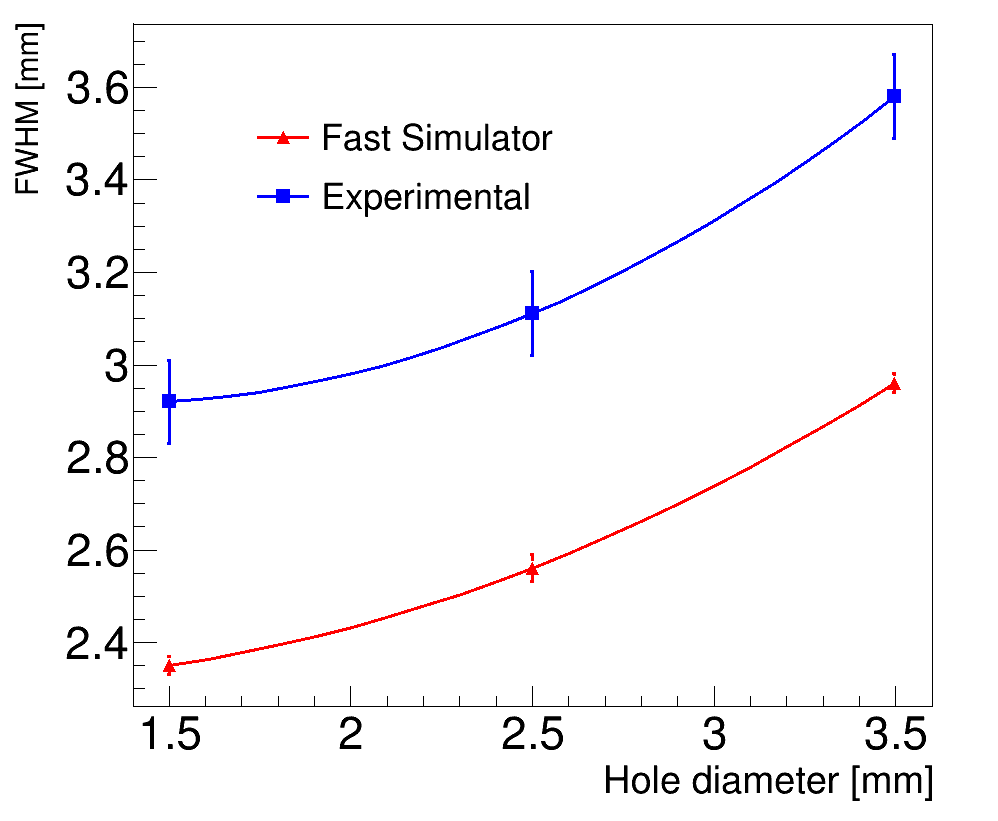}
        \caption{FWHM for 3 diameters.}
        \label{fig:fwhm}
    \end{subfigure}
    \caption{Preliminary results. Fast simulator results in red and experimental results in blue.}
    \end{figure}

A cadmium mask, shown in Fig. \ref{fig:maskCd}, with $\SI{1.5}{\milli\meter}$, $\SI{2.5}{\milli\meter}$ and $\SI{3.5}{\milli\meter}$ hole diameters, was inserted just in front of the detector to obtain the position calibration and estimate the position resolution. We evaluate the fast simulator with these experimental results. In Fig. \ref{fig:chargeDist} we have the comparison of the charge distribution at the readout. An electronic threshold filters signals below, approximately $\SI{100}{\femto\coulomb}$. The projection to one of the axis due to the neutron hit position for the three central holes is shown in Fig. \ref{fig:proj_x}.
    
 Given a set of holes with 3 different diameters we compared the FWHM of the simulated result with the one obtained in the experiment, given in Fig. \ref{fig:fwhm}. The differences between the experimental and fast simulator data can be due to noise, lack of homogeneity in the neutron beam and others effects that were not considered in the simulation and are under investigation.
    
A benchmark of a full and fast simulator shows that the fast simulator is $4$ orders of magnitude faster. Performing the test using an Intel Core i5-8265U CPU @1.60 GHz and 8GB of RAM, the full simulation spends an average of 42 hours while the fast simulator spends an average of 16 seconds for 1 event.

\section{Perspectives}
The next steps in this work consist in a better understanding of the experimental background and electronic noise in order to improve the simulation and to study possible optimization of the detector mainly in terms of position resolution.

\section*{References}
\bibliographystyle{iopart-num}
\bibliography{tipp2021}
\end{document}